# Adaptively Sparse Regularization for Blind Image Restoration

**Ningshan Xu** [1, 2]

[1] Institute of Optics and Electronics, Chinese Academy of Sciences, Chengdu, 610209, China;
[2] The School of Electronic, Electrical and Communication Engineering, University of Chinese Academy of Sciences, Beijing, 100049, China; xuningshan16@mails.ucas.ac.cn;

**Abstract:** Image quality is the basis of image communication and understanding tasks. Due to the blur and noise effects caused by imaging, transmission and other processes, the image quality is degraded. Blind image restoration is widely used to improve image quality, where the main goal is to faithfully estimate the blur kernel and the latent sharp image. In this study, based on experimental observation and research, an adaptively sparse regularized minimization method is originally proposed. The high-order gradients combine with low-order ones to form a hybrid regularization term, and an adaptive operator derived from the image entropy is introduced to maintain a good convergence. Extensive experiments were conducted on different blur kernels and images. Compared with existing state-of-the-art blind deblurring methods, our method demonstrates superiority on the recovery accuracy.

## 1. Introduction

Digital images are common to suffer from degradations caused by imaging systems during formation, storage, transmission, and other processes, leading to image noise, motion blur, Gaussian blur and others. The image quality is then degraded, and the follow-up image understanding and interpretation applications are seriously hampered. Therefore, to assure better application of digital images, image restoration is needed to improve the image quality.

The quality degradation of single-channel image is commonly modeled as,

$$g = f * h + n \tag{1}$$

where $h$ is the point spread function (PSF, *a. k. a* blur kernel) and $n$ is the random additive noise. The image restoration task equals to reconstruct image $f$ from the observed image $g$. Early models to perform image restoration are often under the assumption that the PSF was pre-obtained to allow non-blind deblurring. However, they are prone to ringing artifacts and losing fine image details. Besides, in most practical engineering applications, we have merely little knowledge of the PSF. For such cases, compared with the non-blind method, blind image restoration where both the clear image and blur kernel need to be estimated is necessarily demanded.

*1.1. Related Works*

Over the last few decades, a vast amount of fruitful works have been contributed to establish the image restoration community. The majority of existing blind deblurring methods are based on iterative optimization. Works were developed with exploiting various types of image priors and/or blur kernel priors. Especially in recent years, sparse prior based methods have experienced a renaissance and showed the capability to produce satisfying recovery results.

It was first reported that the gradient distribution of natural images is heavy-tailed [1, 2]. Based on this finding, Fergus *et al.* [3] proposed that the Gaussian mixture model can be used to fit the heavy-tailed distribution. By first estimating the blur kernel, the latent image can be recovered with the RL deconvolution method. Despite the good convergence, the introduction of RL method causes ringing artifacts. Later, Shan *et al.* [4]

proposed a continuous piecewise function to fit the heavy-tailed gradient distributions, and smoothness constraints were applied to suppress the ringing artifacts. Cho and Lee [5] then improved work [4] by performing bilateral filter and edge thresholding in each iteration to remove image noise as well as select significant edges. However, the accuracy of blur kernel estimation is easy to be affected by inconspicuous edges in the image. Levin *et al.* [6] attempted to conform the natural image gradient maps to hyper-laplacian distributions, a simplified version of maximum a posterior (MAP) is proposed to recover the latent image from motion blur. In [7], Xu *et al.* found that the intermediate products of blind image restoration are featured with sparse edge structures. They presented a mathematically sound $L_0$-norm sparse expression which demonstrates immediate energy decreasing and fast speeds for convergence. After that, Wang *et al.* [8] proposed a fused $L_0$-$L_1$ regularization scheme for blind motion deblurring. These sparse prior based approaches have achieved great success with producing sharper recovery results.

Complementary to the iterative-based approaches, the learning-based image deblurring methods have made a rapid development in the last decade. There are many Convolutional Neural Network (CNN) based methods [9-14], which often outperforms existing iterative optimization methods. Zhang *et al.* [12] proposed a network consists of three deep CNNs and a recurrent neural network (RNN). Xu *et al.* [13] adopted CNN to regularize edge enhancement for kernel and image estimation. Li *et al.* [14] integrated neural networks with traditional total-variation regularization algorithms to produce a method with both performance and computational benefits.

Despite the efficiency of learning based methods in determine the parameter set, they share a common disadvantage of depending on the underlying training datasets. Due to the complexity of real-world blurred images, compared with state-of-the-art optimization based methods, most existing learning based methods have limitation on obtaining sufficient datasets for training. What's more, the structures of the networks are often empirically determined and their actual functionality is hard to interpret [14].

*1.2. Our Contributions*

The motivation for our work is twofold. First, according to our experimental investigation and findings, though vast research efforts have been made in the conventional image restoration community, great potentials for blind image restoration still exists (refer to Section 2.1 for more details). Second, there has been a growing trend to tactically combine different image priors or kernel priors, and thus we attempt to develop a more effective prior model to further promote the image restoration performance.

In this work, an adaptively sparse regularized minimization method is originally proposed. Sparsity of high-order derivatives of natural images is studied and a mixed regularization model is introduced. Two main contributions are as follows:
- The low-order and high-order image gradient derivatives are combined to form a hybrid sparse prior model, which presents promising abilities to image detail reconstruction and blur kernel estimation.
- A strategy that can adaptively optimize the iteration process is developed. Derived from image entropy, the introduced adaptive operator leads to a consistent minimization of the objective function, ensuring a better convergence.

Compared with existing state-of-the-art blind deblurring methods, our method demonstrates superiority on restoration performance. We show that our method can generate images with better image quality.

**2. Methodology**

*2.1. Proposed regularzation scheme*

Benefit from its ability of image smoothing, high-order derivatives have been successfully applied to the image restoration method based on Total Variation minimization [15]. However, as we investigated the statistical characteristics of high-order derivatives, we found that the high-order gradient distribution of natural image is even more sparse-featured than the low-order one, as is shown in Figure 1. As is known that the energy minimization problem regarding objective function can benefit a lot from signal spasity, the sparsity of high-order gradient enables its application to such problem. This observation inspired us on applying high-order derivatives to sparse regularization study.

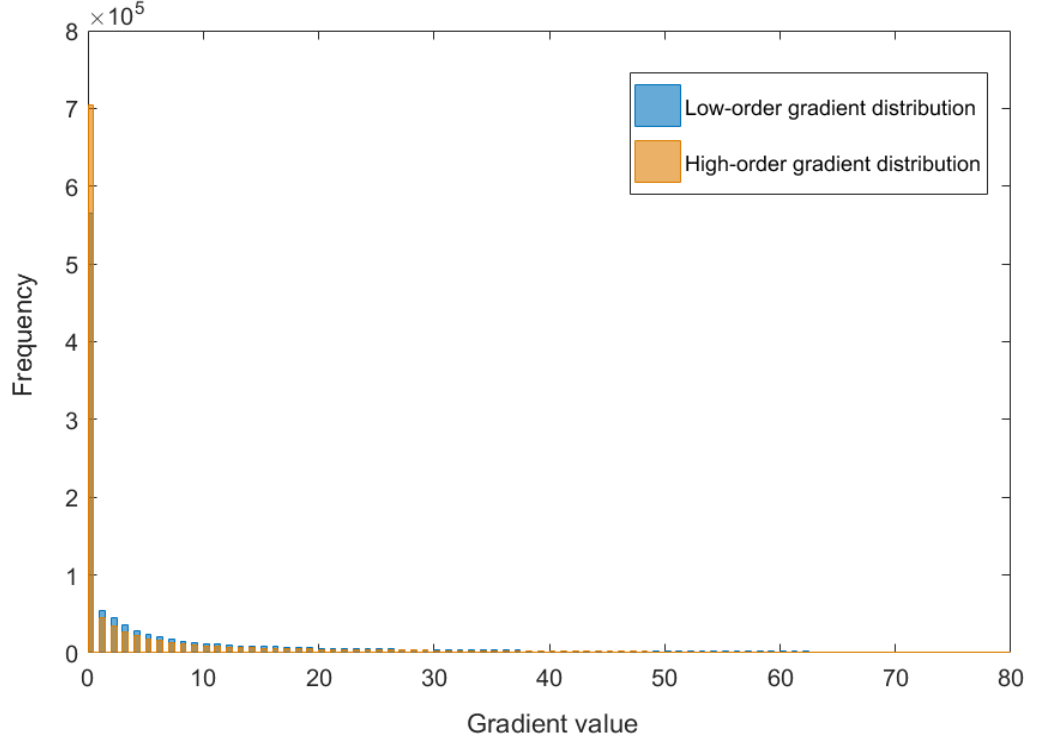

**Figure 1.** An example of distributions of low-order and high-order gradient maps of clear natural images. The orange bar is far over the blue one at the zero position, and we can infer that the high-order gradient distribution shows stronger sparsity. This regularity exists in both horizontal and vertical directions, here only the horizontal gradient distribution is presented for example.

Sparse regularized iterative optimization methods commonly follow the basic model given by (2),

$$(f^*, h^*) = \min_{f,h} \frac{\gamma}{2} \|f * h - g\|^2 + \alpha_f R(f) + \alpha_h R(h) \qquad (2)$$

where $f$ denotes the latent image, $g$ denotes the degraded image and $h$ denotes the blur kernel, * denotes convolution operator. $R(f)$ and $R(h)$ are regularization terms that make for the right solution and away from an infinite number of trivials or other unwanted solutions.

[16] presented that sparse prior well describes the heavy-tailed distribution of the first-order gradients, which helps to produce sharp edges while suppressing noise and ringing artifacts. Hence, in the beginning of our research, we attempted to combine the high-order and low-order gradient derivatives as a new sparse regularizer $R(f)$, so that we can benefit from both forms. The regularization model is given as

$$R(f) = \Phi(D_x f, D_y f, D_{xx} f, D_{xy} f, D_{yx} f, D_{yy} f) = \|\nabla f\|^p + \|\nabla^2 f\|^p \qquad (3)$$

where $D_x$ and $D_y$ are first-order differential operators along horizontal and vertical directions respectively, and $D_{xx}$, $D_{xy}$, $D_{yx}$ and $D_{yy}$ are second-order differential operators. $\nabla f$ represents the distribution of the first-order gradient map and $\nabla^2 f$ represents that of the second-order gradient map; they are given by (4) and (5), respectively,

$$\nabla f = \sqrt{D_x f^2 + D_y f^2} \tag{4}$$

$$\nabla^2 f = \sqrt{D_{xx} f^2 + D_{xy} f^2 + D_{yx} f^2 + D_{yy} f^2} \tag{5}$$

However, such regularization scheme cannot satisfyingly reconstruct the latent image, but tend to generate blurry solutions. To deal with such problem, we introduced a controllable parameter $\omega$ to adjust the ratio of two gradient derived regularization terms, and took observation on how the restoration result varies with $\omega$ tuning. It was interesting that we found the preferable $\omega$ changes with different blur scale. For images with relatively severe blur, higher proportion of $\nabla f$ regularization is needed, and for images with gentle blur, $\nabla^2 f$ regularization weights more. Therefore, considering the variability of the iterative optimization process, we managed to develop an efficient iteration scheme to adaptively control the parameter $\omega$ during each iteration.

In this work, the model of $\omega$ is determined as (6). More details regarding the training and fitting are presented in Section 2.5.

$$\omega = 1 + \frac{Ent^2(f)}{Ent^3(f) + 1} \tag{6}$$

where $Ent$ denotes the image entropy. Then, (2) can be redefined by taking substitution of (3) and (6), and we can get the final objective function

$$(f^*, h^*) = \min_{f,h} \begin{array}{l} \frac{\gamma}{2}\|f * h - g\|^2 + \\ \alpha_f(\|\nabla f\|^p + \omega\|\nabla^2 f\|^p) + \alpha_h R(h) \end{array} \tag{7}$$

To efficiently find the solution, we alternately minimize (7) concerning either $f$ or $h$ while keeping the other constant. Particularly, the above function is a nonconvex optimization problem. Such minimizing problems can be efficiently solved by using the split Bregman method proposed in [17]. The split Bregman iteration is fast and easy to realize. Split Bregman method was used in each minimization sub-problem.

*2.2. Sub-problem with respect to $h$*

This sub-problem is reformed by introducing the auxiliary variable $v_h = h$, and we obtain

$$\min_{h,v_h} \frac{\gamma}{2}\|Fh - g\|^2 + \alpha_h R(v_h) \quad s.t. \ h = v_h \tag{8}$$

where $F$ is a convolution operator constructed from $f$ from the previous iteration. To derive the alternating Split Bregman algorithm, we introduce a quadratic penalty function term $b_h$, which results in the following functional

$$\min_{h,v_h,b_h} \frac{\gamma}{2}\|Fh - g\|^2 + \alpha_h R(v_h) + \frac{\beta_h}{2}\|h - v_h - b_h\|^2 \tag{9}$$

As (9) is differentiable, we derive the optimality condition and set it as zero. Then, the closed-form solution of $h$ can be computed in the Fourier transform domain, and the other variables can be updated via a component-wise strategy. The whole solving process is summarized in Algorithm 1.

For the estimated blur kernel, we apply normalization constraints to the positive kernel values and zero to the negative values, which is mathematically described as

$$h(x,y) = \begin{cases} h, h \geq 0 \\ 0, h < 0 \end{cases}, \ \sum_x \sum_y h(x,y) = 1 \tag{10}$$

---

**Algorithm 1** $h$ update

**Input and initialize:** $v_h^0 = b_h^0 = 0$, loop tolerance *tol*, maximum iterations $N$

**Loop**

$$h^{i+1} = \mathcal{F}^{-1}\left(\frac{\overline{\mathcal{F}(f^i)}\mathcal{F}(g)+\frac{\beta_h}{\gamma}\mathcal{F}(v_h^i+b_h^i)}{\overline{\mathcal{F}(f^i)}\mathcal{F}(f^i)+\frac{\beta_h}{\gamma}I}\right),$$

$$v_h^{i+1} = max\left(h^{i+1} - b_h^i - \frac{\alpha_h}{\beta_h}, 0\right),$$

$$b_h^{i+1} = b_h^i - h^{i+1} + v_h^{i+1},$$

$$i = i + 1,$$

**Until** $\frac{\|h^{i+1}-h^i\|^2}{\|h^{i+1}\|^2} \leq tol$ or $i > N$

**Return** $h^i$

### 2.3. Sub-problem with respect to $f$

Similar to the process of sub-problem $h$, we take an effective replacement to separate the minimization of data term and regularizer, $v_x = D_x f$, $v_y = D_y f$, $v_{xx} = D_{xx} f$, $v_{xy} = D_{xy} f$, $v_{yx} = D_{yx} f$, $v_{yy} = D_{yy} f$. This yields the optimization problem below

$$\min_f \frac{\gamma}{2}\|fH - g\|^2 + \alpha_f \Phi(v_x, v_y, v_{xx}, v_{xy}, v_{yx}, v_{yy})$$

$$s.t.\ v_x = D_x f, v_y = D_y f, v_{xx} = D_{xx} f, v_{xy} = D_{xy} f, v_{yx} = D_{yx} f, v_{yy} = D_{yy} f$$

(11)

where $H$ is a convolutional operator constructed from $h$. Adding the quadratic penalty function terms again and we can obtain the following functional

$$\min_{f,a} \frac{\gamma}{2}\|fH - g\|^2 + \frac{\beta_f}{2}\|D_x f - v_x - a_x\|^2 + \frac{\beta_f}{2}\|D_y f - v_y - a_y\|^2$$
$$+ \frac{\beta_f}{2}\|D_{xx} f - v_{xx} - a_{xx}\|^2 + \frac{\beta_f}{2}\|D_{xy} f - v_{xy} - a_{xy}\|^2 \quad (12)$$
$$+ \frac{\beta_f}{2}\|D_{yx} f - v_{yx} - a_{yx}\|^2 + \frac{\beta_f}{2}\|D_{yy} f - v_{yy} - a_{yy}\|^2$$

As in the previous case, function (12) can be diagonalized by 2D Fourier transform and therefore solved directly. The overall update algorithm for $f$ is as follows.

**Algorithm 2** $f$ update

**Input and initialize:** $v_x^0 = v_y^0 = v_{xx}^0 = v_{xy}^0 = v_{yx}^0 = v_{yy}^0 = a_x^0 = a_y^0 = a_{xx}^0 = a_{xy}^0 = a_{yx}^0 = a_{yy}^0 = 0$, loop tolerance $tol$, maximum iterations $N$

**Loop**

$$f^{i+1} = \mathcal{F}^{-1}\left(\frac{\overline{\mathcal{F}(h^i)}\mathcal{F}(g)+\frac{\beta_f}{\gamma}\left(\begin{array}{c}\overline{\mathcal{F}(D_x)}\mathcal{F}(v_x^i+a_x^i)+\overline{\mathcal{F}(D_y)}\mathcal{F}(v_y^i+a_y^i)+\overline{\mathcal{F}(D_{xx})}\mathcal{F}(v_{xx}^i+a_{xx}^i)\\+\overline{\mathcal{F}(D_{xy})}\mathcal{F}(v_{xy}^i+a_{xy}^i)+\overline{\mathcal{F}(D_{yx})}\mathcal{F}(v_{yx}^i+a_{yx}^i)+\overline{\mathcal{F}(D_{yy})}\mathcal{F}(v_{yy}^i+a_{yy}^i)\end{array}\right)}{\overline{\mathcal{F}(h^i)}\mathcal{F}(h^i)+\frac{\beta_f}{\gamma}\left(\begin{array}{c}\overline{\mathcal{F}(D_x)}\mathcal{F}(D_x)+\overline{\mathcal{F}(D_y)}\mathcal{F}(D_y)+\overline{\mathcal{F}(D_{xx})}\mathcal{F}(D_{xx})+\overline{\mathcal{F}(D_{xy})}\mathcal{F}(D_{xy})\\+\overline{\mathcal{F}(D_{yx})}\mathcal{F}(D_{yx})+\overline{\mathcal{F}(D_{yy})}\mathcal{F}(D_{yy})\end{array}\right)}\right),$$

$$a_x^{i+1} = a_x^i - D_x f^{i+1} + d_x^{i+1},$$

$$a_y^{i+1} = a_y^i - D_y f^{i+1} + d_y^{i+1},$$

$$a_{xx}^{i+1} = a_{xx}^i - D_{xx} f^{i+1} + d_{xx}^{i+1},$$

$$a_{xy}^{i+1} = a_{xy}^i - D_{xy} f^{i+1} + d_{xy}^{i+1},$$

$$a_{yx}^{i+1} = a_{yx}^i - D_{yx} f^{i+1} + d_{yx}^{i+1},$$

$$a_{yy}^{i+1} = a_{yy}^i - D_{yy} f^{i+1} + d_{yy}^{i+1},$$

$$i = i + 1$$

| Until | $\frac{\|f^{i+1}-f^i\|^2}{\|f^{i+1}\|^2} \leq tol$ or $i > N$ |
|---|---|
| Return | $f^i$ |

### 2.4. Sub-problem with respect to $v$

In $f$ update sub-problem, minimization concerning $v$ is a non-convex problem in $L_p$-norm space. The objective function is given as

$$
\begin{aligned}
&\left(v_x^*, v_y^*, v_{xx}^*, v_{xy}^*, v_{yx}^*, v_{yy}^*\right) \\
&= \arg\min \alpha_f \|v_x\|^p + \alpha_f \|v_y\|^p + \alpha_f \omega \|v_{xx}\|^p + \alpha_f \omega \|v_{xy}\|^p \\
&\quad + \alpha_f \omega \|v_{yx}\|^p + \alpha_f \omega \|v_{yy}\|^p + \frac{\beta_f}{2} \|D_x f - v_x - a_x\|^2 \\
&\quad + \frac{\beta_f}{2} \|D_y f - v_y - a_y\|^2 + \frac{\beta_f}{2} \|D_{xx} f - v_{xx} - a_{xx}\|^2 \\
&\quad + \frac{\beta_f}{2} \|D_{xy} f - v_{xy} - a_{xy}\|^2 + \frac{\beta_f}{2} \|D_{yx} f - v_{yx} - a_{yx}\|^2 \\
&\quad + \frac{\beta_f}{2} \|D_{yy} f - v_{yy} - a_{yy}\|^2
\end{aligned} \tag{13}
$$

which is equivalent to solving the minimization $x^* = \arg\min_w \|x\|^p + \frac{\alpha}{2}\|x-w\|^p$. In the case of $0 < p < 1$, no closed-form solution exists of $x$. By using the lookup table (LUT) method [18], we can quickly come to an approximate solution.

### 2.5. Modeling details

For the finding we mentioned about the proportion of $\nabla^2 f$ and $\nabla f$, we proposed to introduce an adaptive parameter to adjust the ratio of two terms. Since it's difficult to define the mathematical representation of the extra introduced parameter, the parameter was modeled by tuning and data fitting.

Considering that the image blur level matters a lot according to our observations, evaluation of image blur is necessary for the adaption purpose. Because of its easy and fast computation, the image entropy is adopted to measure the blur level and then for data fitting. 32 images in total from the dataset [6] was adopted for parameter tuning. Only $\omega$ is modified while keeping the other constant. For each blurred image, $\omega$ starts with 1 and is tuned on the basis of experimental observation, the SSIM between recovery result and the ground truth is obtained after each tuning process. The tuning criterion is that the parameter value leading to a higher SSIM is chosen. After the parameter tuning, 32 data points of the form (Entropy, $\omega$) are used for data fitting, and the final representation is determined as (6).

Through analysis of (6), we propose a possible explanation on how $\omega$ affects the regularization process. Specifically, experiments show that with the image blur level increases, the high-order gradient distribution is less heavy-tailed, while the low-order distribution is relatively mildly affected. As shown in Figure 2. Under such circumstance, the high-order derivatives cannot be well described by the hyper-laplacian distribution, and errors caused by modeling are correspondingly introduced to the restoration process. Therefore, to protect the iterative optimization from modeling error, the proportion of high-order regularization term needs to be decreased at high blur level, and thus the functional mechanism of the adaptive parameter is explained.

Note that with parameter tuning and data fitting, errors are certainly embedded in. The significance of the proposed adaptive strategy lies in its promising performance over different image dataset and comparison with other methods. In our future work, the proposed assumption needs to be verified and further investigation regarding behavior of high-order sparse regularization is necessary.

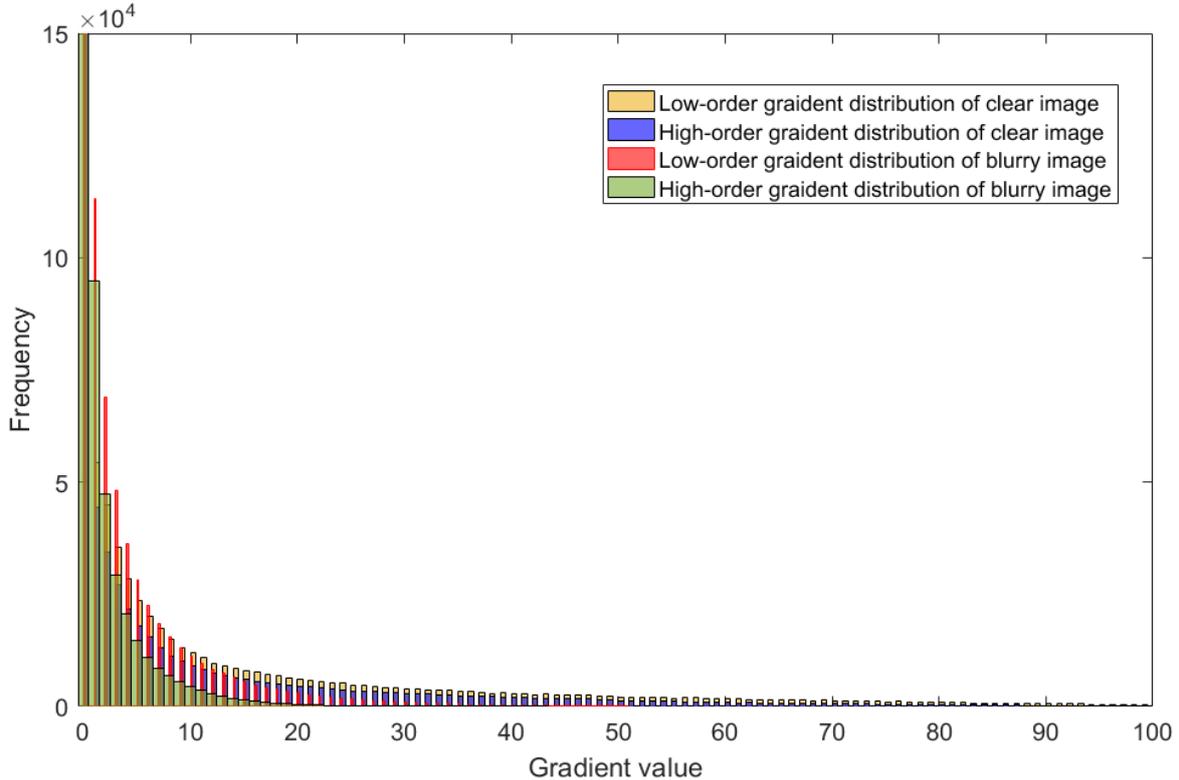

**Figure 1.** An example of the variations of image gradients from clear to blurry (Please zoom in for a better view). Compared with the blue bar, the green bar is obviously less heavy-tailed, while the red bar has a lower rate of change.

*2.6. Implementation setups*

In addition to the aforementioned details, the proposed model was developed in a coarse-to-fine fashion. We multiply the weight of the fidelity term $\gamma$ by 1.5 after each iteration, that is, gradually release the constraints imposed by regularization, which is conductive to reduce the piecewise smoothing effects on the recovery image [19]. Other parameters are set as $\gamma = 1, \alpha_f = 0.01, \alpha_h = 10, \beta_f = 1, \beta_h = 10^4, p = 0.3, N = 10$. Besides, to prevent the algorithm from falling into local minima, we set the stopping criterion for iteration as $tol = 0.001$.

**3. Experiments and analysis**

In this section, we analyze the restoration ability of our algorithm against existing state-of-the-art methods. Two image datasets were employed for experiment. The BSDS dataset [14] is a synthetic one, it adopts 200 images from the test portion of the BSDS500 dataset [20] and 4 manually generated test kernels to synthesize 800 blurred images. Gaussian noise of standard deviation 0.01 is added. The GOPRO dataset [9] is a real-life one, blurry in this dataset is caused by lens imperfections and relative motion.

The algorithm performances are evaluated with two acknowledged metrics; the Structural Similarity index (SSIM) and the Peak-Signal-to-Noise Ratio (PSNR) metric. Six outstanding deblurring methods are considered for comparison, including Krishnan [21], Xu and Jia [7], Kotera [16], Hosseini [22], Li [14] and Nah [9]. The former four are iterative-based methods and the latter two are learning-based methods.

*3.1. Ablation experiment*

First, to demonstrate the effectiveness of our proposed regularization scheme, the ablation experiment was performed. Images used for experiment are with the latter half of the BSDS dataset. The average PSNR and SSIM values are presented in Table 1 and Figure 3 shows an example of ablation experiment results. In the light of the results, we can clearly infer that with measures of the high-order derivatives and the adaptive parameter ω, the restored images have better visual effect, and the restoration accuracy is significantly improved against the baseline. Hence, the proposed hybrid regularization model demonstrates its ability to image details estimation.

**Table 1.** The average performance of ablation study regarding the proposed regularization scheme.

| Measure | Blurred | $\|\nabla f\|^p$ | $\|\nabla f\|^p + \|\nabla^2 f\|^p$ | $\|\nabla f\|^p + \omega\|\nabla^2 f\|^p$ |
|---|---|---|---|---|
| SSIM | 0.5436 | 0.7110 | 0.7407 | 0.8280 |
| PSNR | 25.1436 | 26.0087 | 26.6255 | 26.7176 |

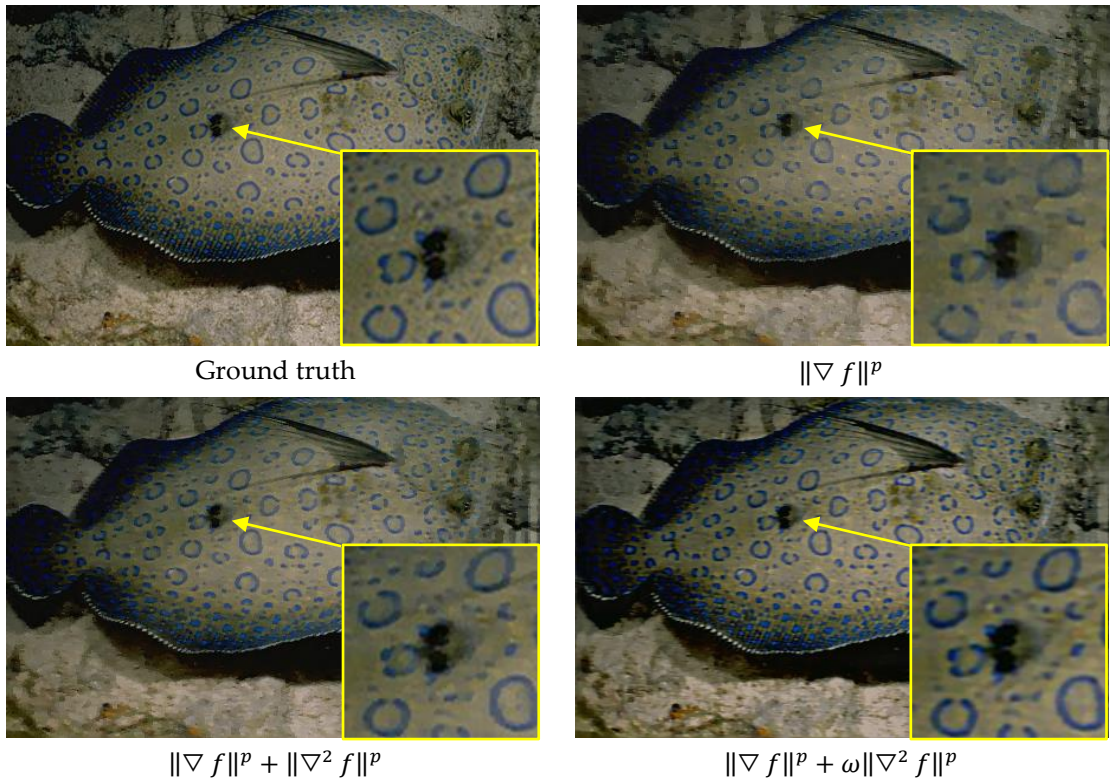

Ground truth     $\|\nabla f\|^p$

$\|\nabla f\|^p + \|\nabla^2 f\|^p$     $\|\nabla f\|^p + \omega\|\nabla^2 f\|^p$

**Figure 3.** An example of ablation experiment of the proposed regularization scheme.

### 3.2. Experiments of synthetic data

We compare our method with existing blind deblurring methods on the BSDS dataset. The SSIM and PSNR values were obtained and the overall average results are presented in Table 2. An example of the ground truth image and the deblurred results is shown in Figure 4.

From Table 2 and Figure 4 we can recognize that over different test data, our method performs restoration at higher accuracy and therefore the estimated images are more faithful to the ground truth. Additionally, when examine the example presented in Figure 4, we can observe that our method raises less extra artifacts.

**Table 2.** Average deconvolution performance on dataset [20]. The best performed ones are marked in bold.

| Metrics | Li [14] | Nah [9] | Krishnan [21] | Xu and Jia [7] | Hossein [22] | Ours |
|---|---|---|---|---|---|---|
| SSIM | 0.8303 | 0.8257 | 0.7270 | 0.7841 | 0.5404 | **0.8403** |
| PSNR | 20.8716 | 25.3978 | 22.8246 | 24.4006 | 20.7021 | **25.9269** |

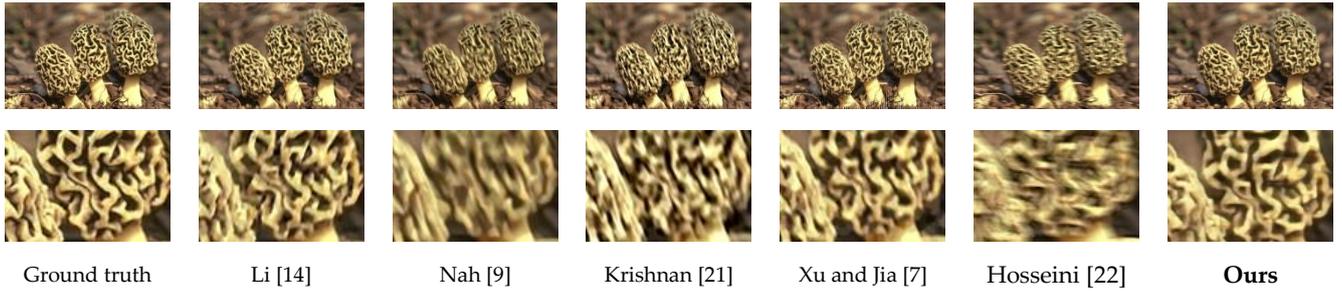

**Figure 4.** Examples of qualitative comparisons on the BSDS dataset [20]. Columns from left to right: blurry image, Li, Nah, Krishnan, Xu and Jia, Hosseini, and ours.

### 3.3. Experiments of real-life images

To further prove the ability of proposed algorithm to practical applications, we tested our method on real distorted images dataset [9]. Figure 5 presents an example of the estimated images in comparison. Table 3 reports the overall averaged SSIM and PSNR values. It can be seen that our method produces sharper edges and smoother details. However, due to the unavoidably inexact kernel estimation, the ringing artifacts and the blurry effects still exist, which needs to be investigated in our further work.

**Table 3.** Average deconvolution performance on dataset [9]. The best performed ones are marked in bold.

| Metrics | Li [14] | Nah [9] | Krishnan [21] | Xu and Jia [7] | Hossein [22] | Ours |
| --- | --- | --- | --- | --- | --- | --- |
| SSIM | 0.8419 | **0.8752** | 0.7901 | 0.8147 | 0.8152 | 0.8723 |
| PSNR | 27.5433 | 28.2621 | 25.0031 | 26.6808 | 27.5458 | **30.4183** |

**Figure 5.** An example of real-life blurry image from the GOPRO dataset [9]. Columns from left to right: blurry image, Li, Nah, Krishnan, Xu and Jia, Hosseini, and ours.

### 4. Conclusion

In this study, we present a blind image restoration method with an adaptively hybrid sparse regularization scheme. The quantitative evaluations and comparisons demonstrate the superiority on recovering image with both smoothness and sharpness of our method. Extensive experiments on both synthetic and real-life data prove that our method outperforms existing state-of-the-art methods in terms of restoration accuracy.

According to our current work, the sparsity of high-order derivatives has been well applied to the sparse regularization method. In our future work, the proposed assump-tion (refer to Section 2.5) needs to be verified and further investigation regarding behavior of high-order regularization term is necessary.

**Conflicts of Interest:** The authors declare no conflict of interest.